\documentclass[a4paper,10pt,twoside]{cpc-hepnp}

\usepackage{multicol}
\usepackage{graphicx}
\usepackage{booktabs}
\usepackage{amssymb,bm,mathrsfs,bbm,amscd}
\usepackage[tbtags]{amsmath}
\usepackage{lastpage}

\begin{document}

\fancyhead[co]{\footnotesize A. Gal: $\bar K$ nuclear dynamics}  

\title{$\bar K$-nucleus dynamics: from quasibound states \\ 
to kaon condensation}

\author{Avraham Gal \email{avragal@vms.huji.ac.il}} 

\maketitle 

\address{Racah Institute of Physics, The Hebrew University,
Jerusalem~91904, Israel \\ }

\begin{abstract}
Coupled-channel $\bar K N$ dynamics near threshold and its repercussions in 
few-body $\bar K$-nuclear systems are briefly reviewed, highlighting studies 
of a $K^-pp$ quasibound state. In heavier nuclei, the extension of mean-field 
calculations to multi-$\bar K$ nuclear and hypernuclear quasibound states is 
discussed. It is concluded that strangeness in finite self-bound systems is 
realized through hyperons, with no room for kaon condensation. 
\end{abstract} 



\begin{multicols}{2}

\section{Introduction}
\label{intro}

The gross features of low-energy $\bar K N$ physics are encapsulated 
in the leading-order Tomozawa-Weinberg (TW) vector term of the chiral 
effective Lagrangian~\cite{TW66}. The Born approximation for the 
$\bar K$-{\it nuclear} optical potential $V_{\bar K}$ due to the TW 
interaction term yields then a sizable attraction: 
\begin{equation} 
\label{eq:chiral} 
V_{\bar K}=-\frac{3}{8f_{\pi}^2}~\rho\sim -55~\frac{\rho}{\rho_0}~~{\rm MeV} 
\end{equation} 
for $\rho _0 = 0.16$ fm$^{-3}$, where $f_{\pi} \sim 93$ MeV is the 
pseudoscalar meson decay constant. Iterating the TW term plus the less 
significant NLO terms, within an {\it in-medium} coupled-channel approach 
constrained by the $\bar K N - \pi \Sigma - \pi \Lambda$ data near the 
$\bar K N$ threshold, roughly doubles this $\bar K$-nucleus attraction. 
A major uncertainty in these chirally based studies arises from fitting 
the $\Lambda(1405)$ resonance by the imaginary part of the $\pi\Sigma(I=0)$ 
amplitude calculated within the same coupled channels chiral scheme. 
However, irrespective of this uncertainty, the $\Lambda(1405)$ which may be 
viewed as a $K^-p$ quasibound state quickly dissolves in the nuclear medium 
at low density, so that the repulsive free-space scattering length $a_{K^-p}$, 
as function of $\rho$, becomes {\it attractive} well below $\rho _0$. 
Adding the weakly density dependent  $I=1$ attractive scattering length 
$a_{K^-n}$, the resulting in-medium $\bar K N$ isoscalar scattering length 
$b_0(\rho)={\frac{1}{2}}(a_{K^-p}(\rho)+a_{K^-n}(\rho)$) translates into 
a strongly attractive $V_{\bar K}$: 
\begin{equation} 
\label{eq:trho} 
{\rm Re}V_{\bar K}(\rho_0)=-{\frac{2\pi}{\mu_{KN}}}{\rm Re}b_0(\rho_0)\rho_0 
\sim -110~{\rm MeV}\,. 
\end{equation} 
The underlying $K^-p$ forward scattering amplitude is shown in Fig.~\ref{fig1}. 

\begin{center} 
\includegraphics[width=7cm]{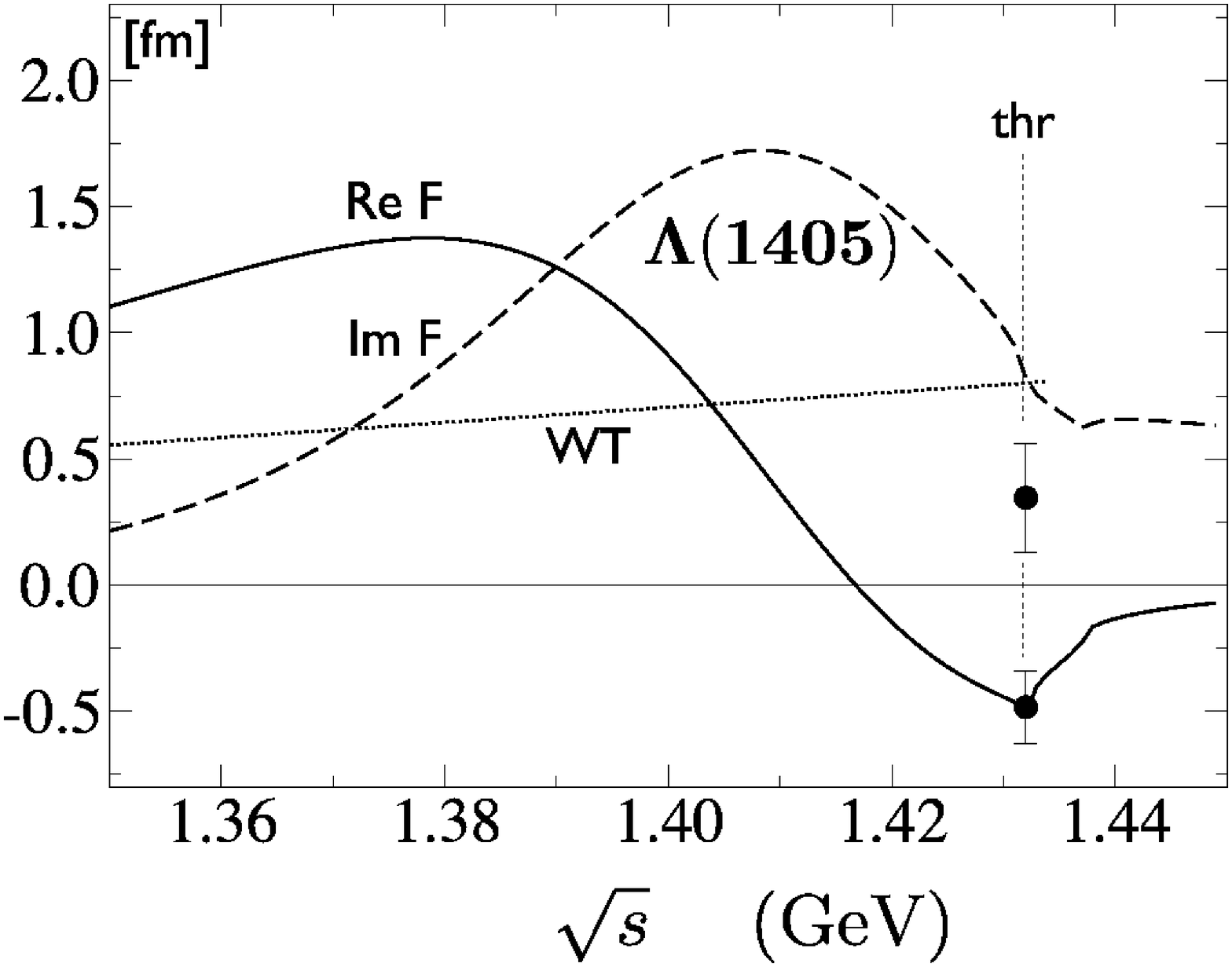} 
\figcaption{\label{fig1} The $K^-p$ forward scattering amplitude calculated 
in the chiral $SU(3)$ coupled channel approach \citep{BNW05}. The scattering 
length deduced from the DEAR kaonic hydrogen measurement \citep{DEAR} is also 
shown. The dotted, almost horizontal line indicates the TW amplitude. 
Figure adapted from Ref.~\citep{WH08}.} 
\end{center} 

Comprehensive fits to the strong-interaction shifts and widths of $K^-$-atom 
levels provide phenomenological evidence for a strongly attractive, 
and also strongly absorptive $\bar K$-nucleus interaction near 
threshold~\cite{BFG97,FG07}. These fits yield extremely deep density dependent 
optical potentials with nuclear-matter depth $-{\rm Re}V_{\bar K}(\rho_0)\sim 
(150-200)$ MeV at threshold.  
Figure \ref{fig2} illustrates the real part of the best-fit $\bar K$-nucleus 
potential for $^{58}$Ni as obtained for several models. The corresponding 
values of $\chi ^2$ for 65 $K^-$-atom data points are given in parentheses. 
A model-independent Fourier-Bessel (FB) fit \cite{BF07} is also shown, within 
an error band. Just three terms in the FB series, added to a $t\rho $ 
potential, suffice to achieve a $\chi ^2$ as low as 84 and to make the 
potential extremely deep, in agreement with the density-dependent best-fit 
potentials DD and F. In particular, potential F provides by far the best fit 
ever reported for any global $K^-$-atom data fit~\cite{MFG06}, and the lowest 
$\chi ^2$ value as reached by the FB method. 

\begin{center}
\includegraphics[width=7cm]{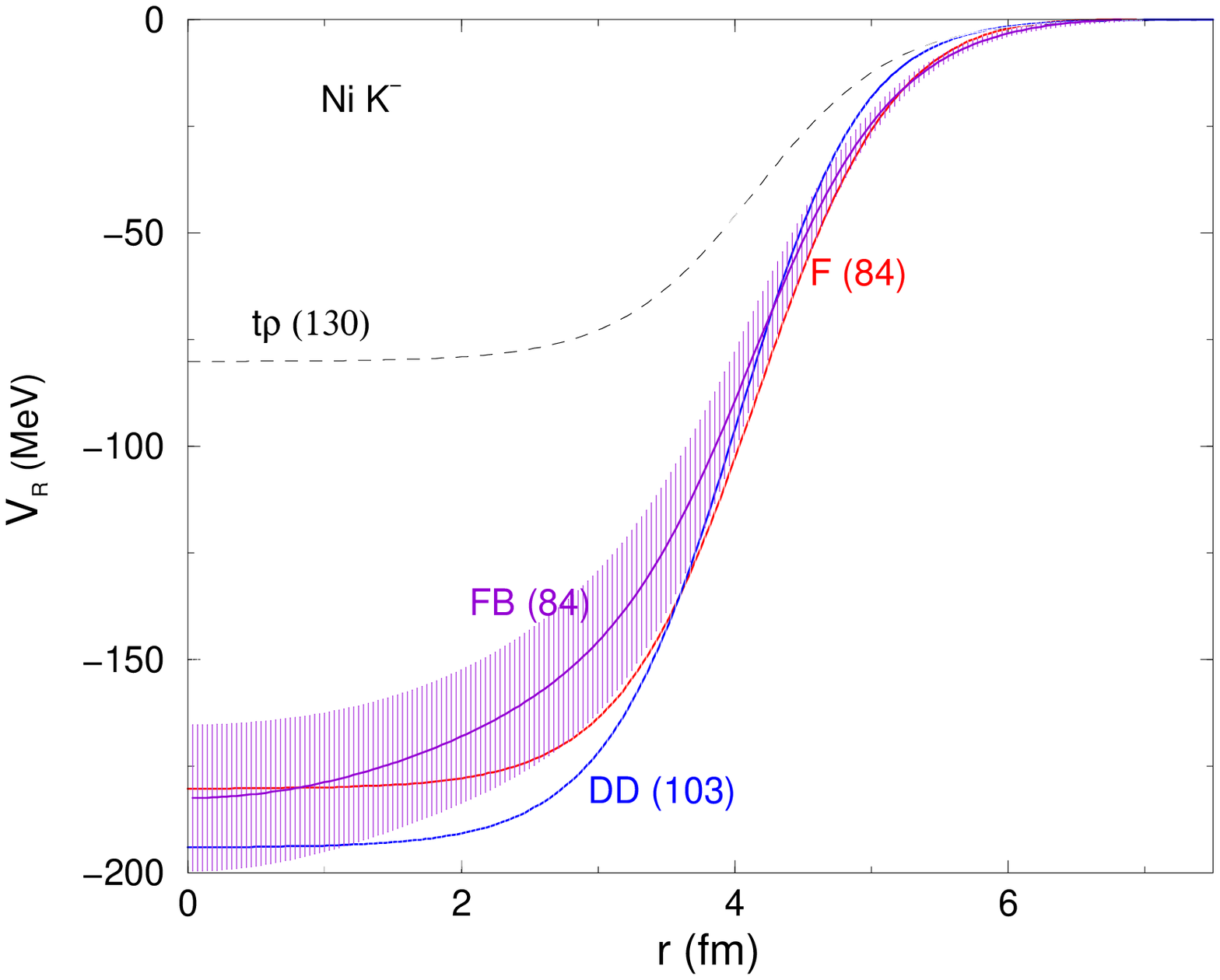}
\figcaption{\label{fig2} Real part of the $\bar K - {^{58}{\rm Ni}}$ 
potential for several density dependent potentials (DD, FB, F) and a $t\rho$ 
potential fitted to kaonic-atom data \citep{FG07}. $\chi ^2$ values are given 
in parentheses.} 
\end{center} 

\begin{center} 
\includegraphics[width=7cm]{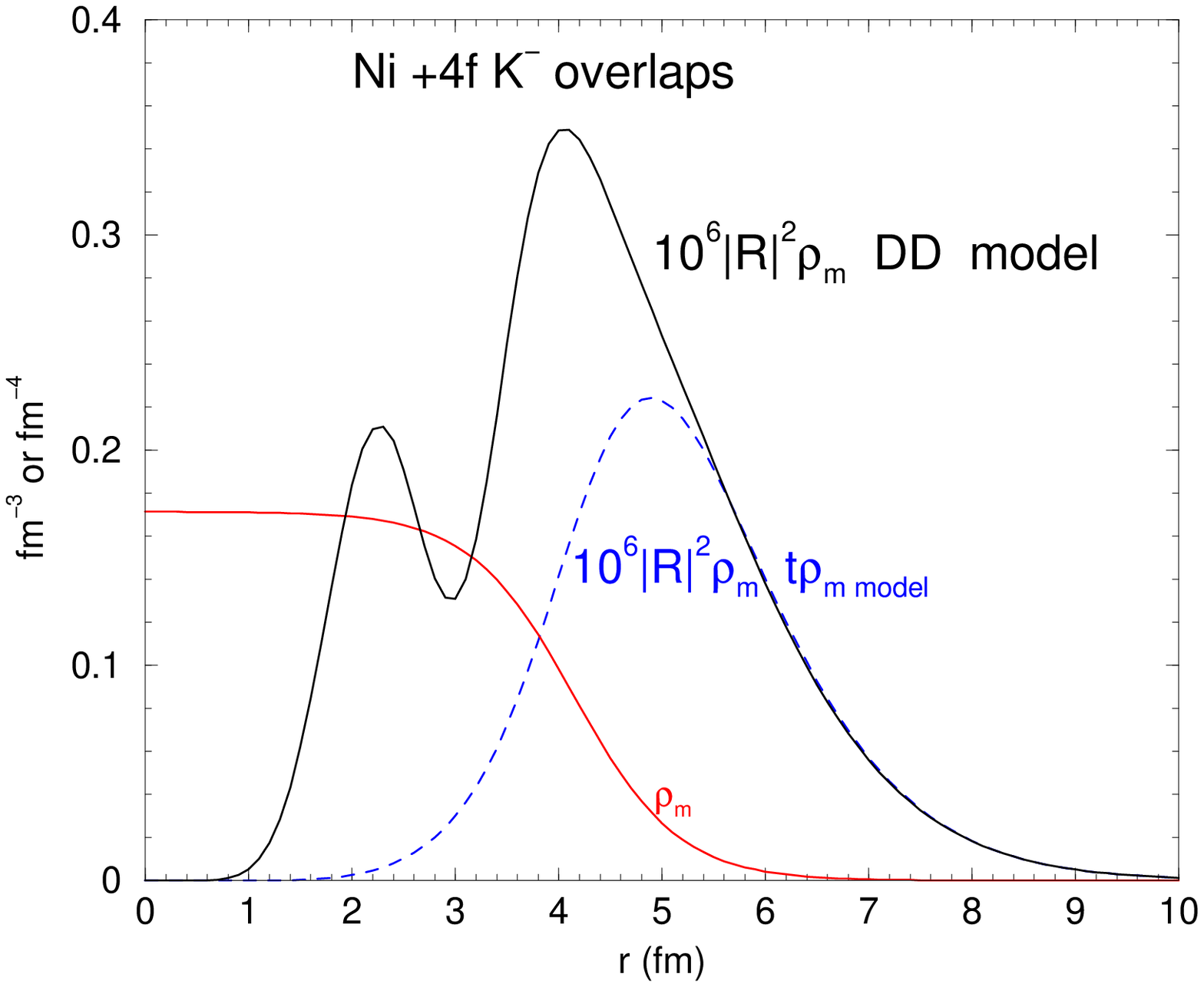} 
\figcaption{\label{fig3} Overlaps of $K^-$ $4f$ atomic radial wavefunctions 
$R$ squared with $\rho_m$ of $^{58}$Ni, for the $t\rho$ and DD fits of 
Fig.~\ref{fig2}. The nuclear matter density $\rho_m$ is shown for 
comparison. Figure provided by Eli Friedman.} 
\end{center} 

In Fig.~\ref{fig3} I show the overlap of the $4f$ atomic radial wavefunction 
squared with the matter density $\rho_m$ in $^{58}$Ni for two of the models 
exhibited in Fig.~\ref{fig2}. The $4f$ atomic orbit is the last circular $K^-$ 
atomic orbit from which the $K^-$ meson undergoes nuclear absorption. 
The figure demonstrates that, whereas this overlap for the shallower $t\rho$ 
potential peaks at nuclear density of order $10\%$ of $\rho_0$, it peaks at 
about $60\%$ of $\rho_0$ for the deeper DD potential and has a secondary peak 
well inside the nucleus. The double-peak structure indicates the existence of 
a $K^-$ strong-interaction $\ell=3$ quasibound state for the DD potential. 
It is clear that whereas within the $t\rho$ potential there is no sensitivity 
to the interior of the nucleus, the opposite holds for the density dependent F 
potential which accesses regions of full nuclear density. This owes partly 
to the smaller imaginary part of F. 

\begin{center} 
\includegraphics[width=7cm]{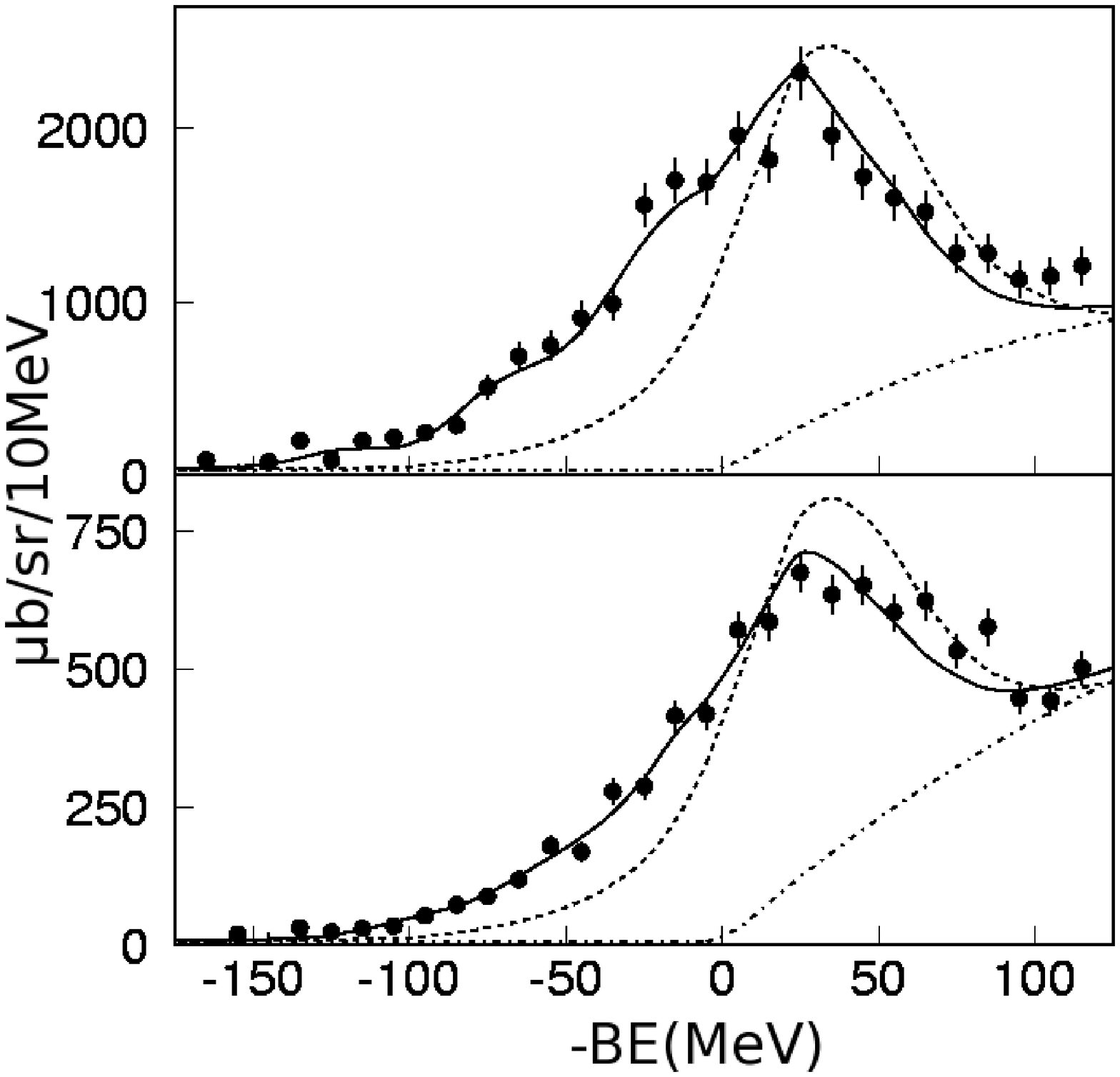} 
\figcaption{\label{fig4} KEK-PS E548 missing mass $(K^-,n)$ (upper) \& 
$(K^-,p)$ (lower) spectra on $^{12}$C at $p_{K^-}=1$~GeV/c \citep{kish07}.} 
\end{center}
 
A fairly new and independent evidence in favor of extremely deep 
$\bar K$-nucleus potentials is provided by $(K^-,n)$ and $(K^-,p)$ 
spectra taken at KEK on $^{12}$C \cite{kish07} and very recently also 
on $^{16}$O (presented in PANIC08) at $p_{K^-}=1$ GeV/c. The $^{12}$C 
spectra are shown in Fig.~\ref{fig4}, where the solid lines represent 
calculations (outlined in Ref.~\cite{YNH06}) 
using potential depths in the range 160-190 MeV. The dashed lines 
correspond to using relatively shallow potentials of depth about 60 MeV 
which may be considered excluded by these data. However, Magas {\it et al.} 
have recently expressed concerns about protons of reactions other than 
those {\it directly} emanating in the $(K^-,p)$ reaction and which could 
explain part of the bound-state region of the measured spectrum without 
invoking a very deep $\bar K$-nuclear potential~\cite{MYS09}. A sufficiently 
deep potential would allow quasibound states bound by over 100 MeV, 
for which the major $\bar K N \to \pi \Sigma$ decay channel is blocked, 
resulting in relatively narrow $\bar K$-nuclear states. Of course, a fairly 
sizable extrapolation is involved in this case using an energy-independent 
potential determined largely near threshold. 

A third class, of shallower 
potentials with Re$V_{\bar K}(\rho_0) \sim -$(40-60) MeV, was obtained by 
imposing a Watson-like self-consistency requirement on the nuclear-medium 
$\bar K N$ $t(\rho)$ matrix that enters the optical potential 
$V_{\bar K}=t(\rho)\rho$~\cite{RO00,CFG01}. This is due to the suppressive 
effect of Im$t(\rho)$ in the $K^-$ propagator of the Lippmann-Schwinger 
equation for $t(\rho)$: 
\begin{equation}\label{eq:sc} 
t(\rho)=v+v\frac{1}{E-H^{(0)}_{\rm mB}-t(\rho)\rho-V_N+{\rm i}0}t ~. 
\end{equation} 
Here $v$ and $t(\rho)$ are coupled-channel meson-baryon (mB) potential and 
{\it in-medium} $t$ matrix, respectively, and $H^{(0)}$ is the kinetic energy 
operator which depends implicitly on the density $\rho$ through the imposition 
of the Pauli principle in $\bar K N$ intermediate states. The $K^-$ optical 
potential $t(\rho)\rho$ and the nucleon potential $V_N$ act only in $\bar K N$ 
intermediate states. A sizable Im$t(\rho)$ leads to exponential decay of 
the propagator $(E-H^{(0)}_{\rm mB}-t(\rho)\rho-V_N+{\rm i}0)^{-1}$, so that 
$t(\rho) \approx v$ thus losing the cooperative coupling effect to the $\pi Y$ 
channels in higher-order terms of $v$. However, one needs then to worry 
about higher orders in the chiral expansion which are not yet in. 

I start this review by making brief remarks on the $\bar KN - \pi\Sigma$ 
system, followed by reviewing topics related to $\bar K$ nuclear quasibound 
states: (i) the $K^-pp$ system as a prototype of few-nucleon quasibound 
states of $\bar K$ mesons; and (ii) multi-$\bar K$ nucleus quasibound states.

\section{$\bar KN - \pi\Sigma$ coupled channels} 
\label{poles} 

\begin{center}
\includegraphics[width=7cm]{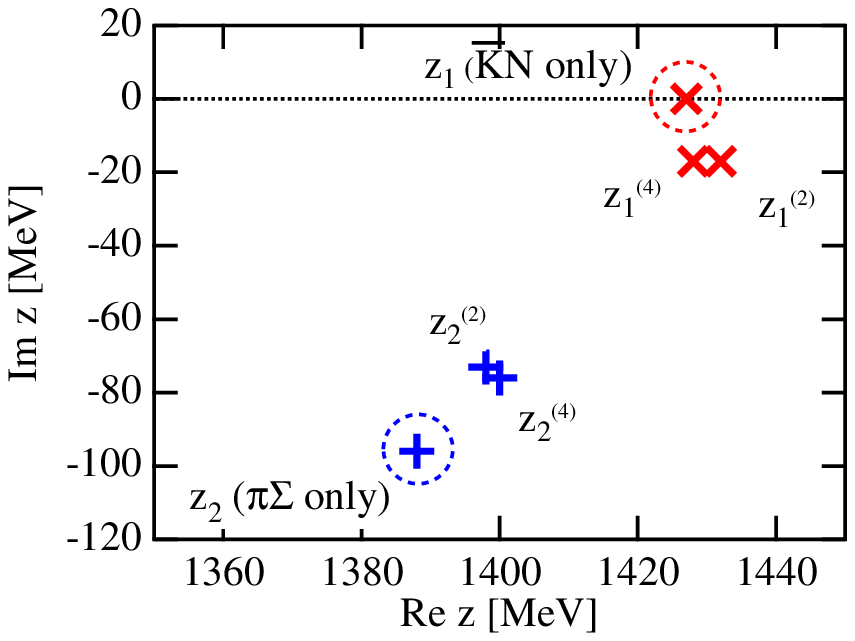}
\figcaption{\label{fig5} Pole positions in the complex energy (z) plane of 
the $\bar K N~(I=0)$ scattering amplitude resulting from single-channel, 
two-channel and full (four-channel) models \citep{HW08}.} 
\end{center} 

\begin{center}
\includegraphics[width=7cm]{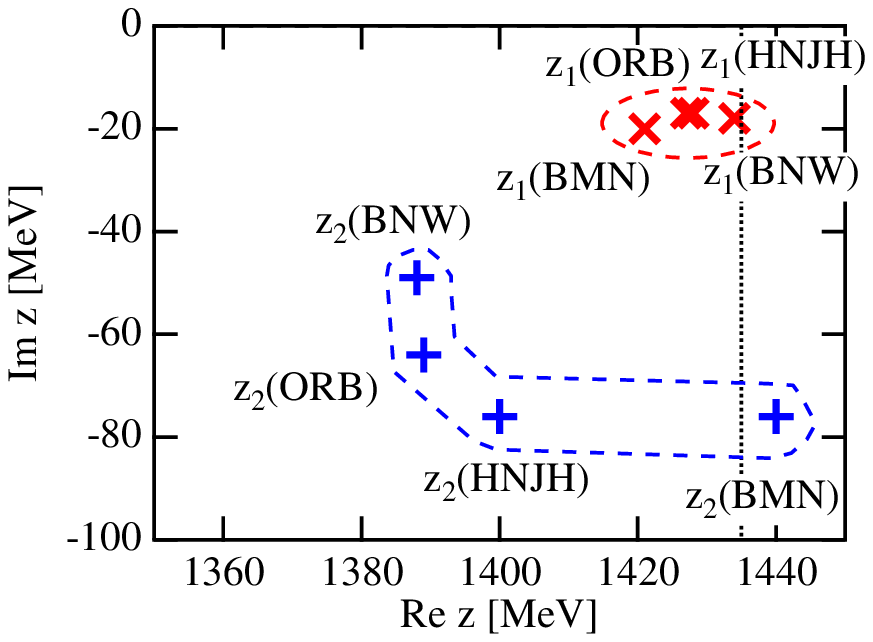} 
\figcaption{\label{fig6} Poles of the $\bar K N~(I=0)$ scattering amplitude 
in several chiral model applications listed in Ref.~\citep{HW08}. The dotted 
vertical line denotes the $\bar K N$ threshold.}
\end{center} 

Modern chirally motivated $\bar KN - \pi\Sigma$ coupled-channel models 
give rise to {\it two} Gamow poles that dominate low-energy $\bar K N$ 
dynamics. Representative pole positions are shown in Fig.~\ref{fig5} and 
their model dependence is demonstrated in Fig.~\ref{fig6}. The $\Lambda(1405)$ 
resonance, studied in final-state $\pi\Sigma$ interactions, is determined in 
these models primarily by the lower pole. This identification is further 
supported, as shown in Fig.~\ref{fig7}, by the trajectory of the lower pole 
which merges into a genuinely $I=0$ bound state below the $\pi\Sigma$ 
threshold when the $\bar K N$ interaction is sufficiently increased. The 
upper pole appears in this chiral model~\cite{CS07} above the $\bar K N$ 
threshold. Its position and the trajectory it follows away from the real 
energy axis make it largely model dependent and sensitive to off-shell 
effects. 

\begin{center}
\includegraphics[width=7cm]{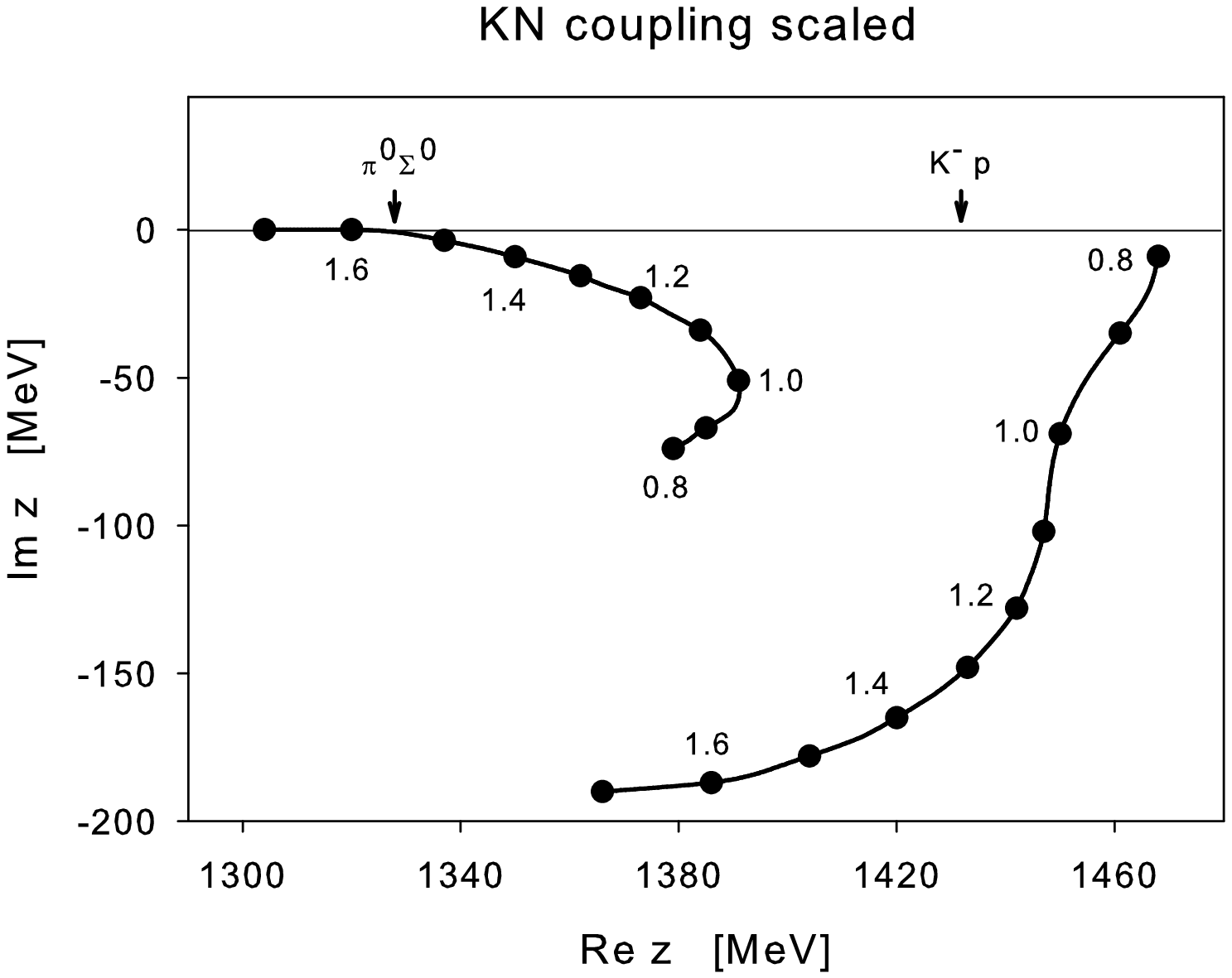} 
\figcaption{\label{fig7} Trajectories of Gamow poles in the complex energy 
(z) plane, on the Riemann sheet [$\Im k_{\bar K N},\Im k_{\pi \Sigma}$] = 
[$+,-$], upon scaling the $\bar K N$ interaction strengths \citep{CG08}. 
The $\pi^0\Sigma^0$ and $K^-p$ thresholds are marked by arrows.} 
\end{center}  

Surprisingly, however, it is the upper two-body pole that is found to affect 
significantly three-body $[\bar K (NN)_{I=1} - \pi\Sigma N]_{I=1/2}$ coupled 
channel calculations for the $K^-pp$ system. 
The trajectory of the {\it only} Gamow pole in this system that qualifies 
for representing a quasibound state is depicted in circles in Fig.~\ref{fig8}. 
Similarly to the trajectory of the $\Lambda(1405)$ in the two-body case, 
this quasibound $\bar K NN$ pole also merges into a genuinely bound state 
below the $\pi\Sigma N$ threshold which becomes, upon extending the model 
space, a quasibound $\pi\Sigma N$ state decaying to the $\pi\Lambda N$ and 
$YN$ lower channels ignored here.

\begin{center}
\includegraphics[width=7cm]{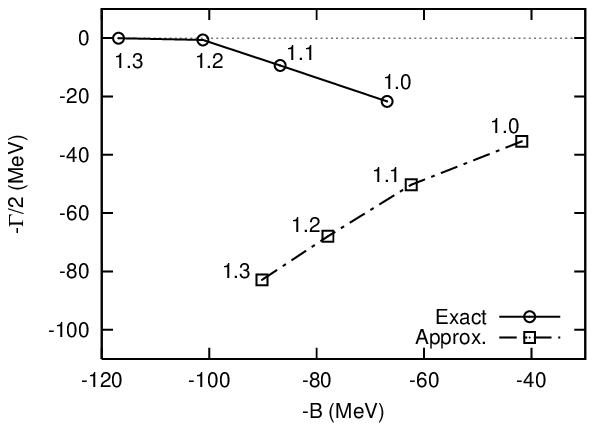} 
\figcaption{\label{fig8} Variation of $\bar K NN(I=1/2)$ quasibound state 
energy with $\bar K N$ interaction strength in three-body coupled-channel 
(circles) and single-channel (squares) calculations \citep{IS08}.} 
\end{center} 
 
\section{Few-nucleon $\bar K$ systems} 
\label{KNN}

The lightest $\bar K$ nuclear configuration maximizing the strongly attractive 
$I=0~\bar K N$ interaction is $[\bar K (NN)_{I=1}]_{I=1/2,J^{\pi}=0^-}$, 
loosely denoted as $K^-pp$. The FINUDA collaboration presented evidence in 
$K^-$ stopped reactions on several nuclear targets for the process $K^-pp \to 
\Lambda p$, interpreting the observed signal as due to a $K^-pp$ deeply bound 
state with $(B, \Gamma) \approx (115, 67)$~MeV \cite{ABB05}. 
However, this interpretation has been challenged in Refs.~\cite{MFG06,MOR06}. 
A new analysis of DISTO $pp \to K^+ \Lambda p$ data claims a $K^-pp$ signal 
with $(B, \Gamma)\approx (103, 118)$~MeV \cite{Yam08}, see Fig.~\ref{fig9}. 
Its location practically on top of the $\pi\Sigma N$ threshold, 
and particularly the large width, are at odds with any of the few-body 
calculations listed below, posing a problem for a $K^-pp$ quasibound state 
interpretation. 

\begin{center}
\includegraphics[width=7cm]{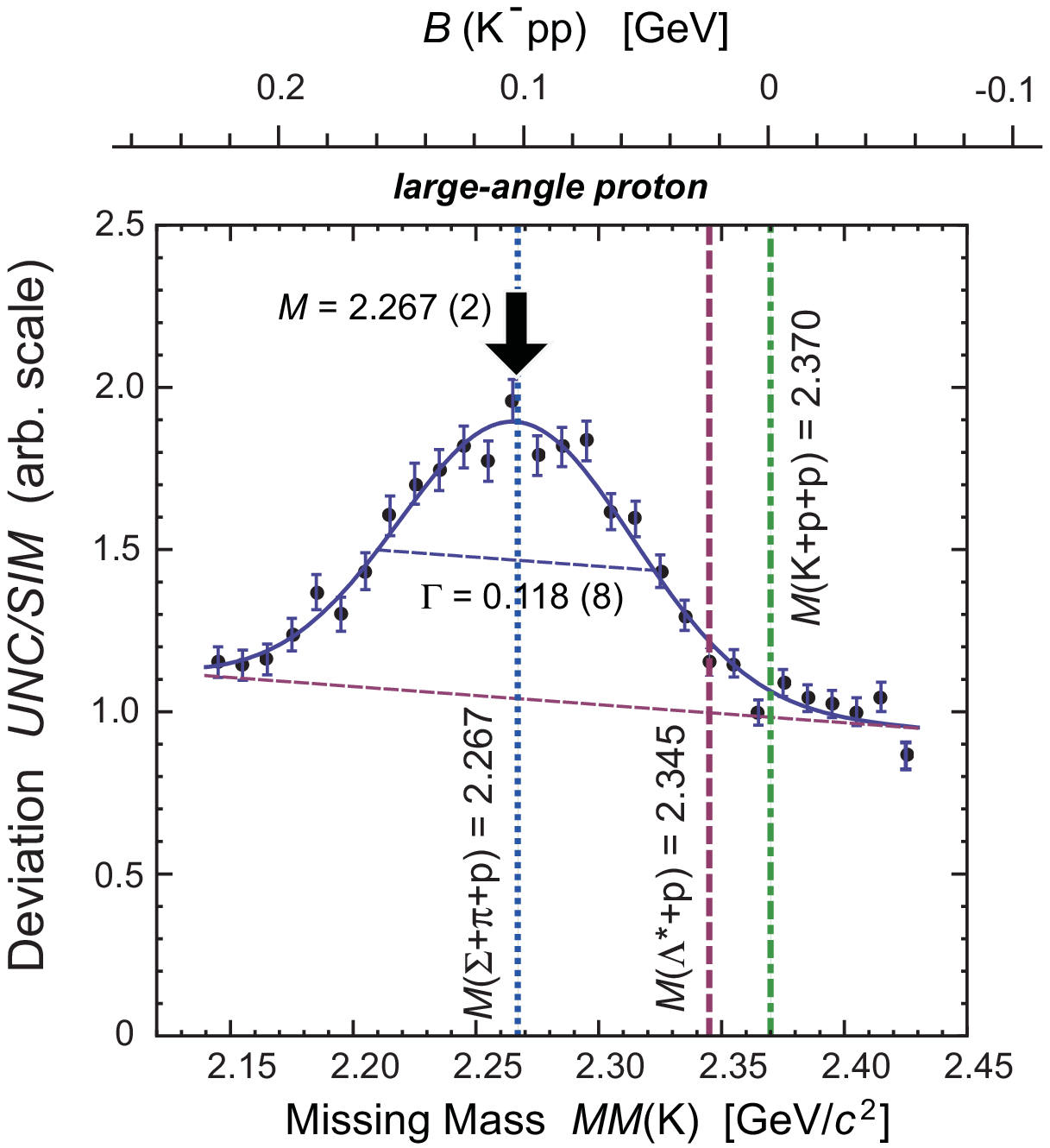} 
\figcaption{\label{fig9} $K^+$ missing-mass spectrum in the reaction 
$pp\to K^+\Lambda p$ measured at $T_p=2.85$~GeV by the DISTO 
Collaboration \citep{Yam08}. The peak structure with a background (thin line) 
gives $M=2267 \pm 2,~\Gamma=118 \pm 8$ MeV.} 
\end{center} 

\end{multicols} 

\begin{center} 
\tabcaption{ \label{tab1}  Calculated $K^-pp$ binding energies ($B_{K^-pp}$), 
mesonic ($\Gamma_{\rm m}$) \& nonmesonic ($\Gamma_{\rm nm}$) widths (in MeV).} 
\vspace{-3mm}
\footnotesize 
\begin{tabular*} {170mm}{@{\extracolsep{\fill}}ccccccc}  
\toprule 
&\multicolumn{2}{c}{$\bar K NN$ single channel} 
&\multicolumn{3}{c}{$\bar K NN - \pi\Sigma N$ coupled channels} \\ 
& variational~\cite{YA02,AY02} & variational~\cite{DHW08} & 
Faddeev~\cite{SGM07} & Faddeev~\cite{IS07} & variational~\cite{WG08} \\ 
\hline
$B_{K^-pp}$ & ~~~48 & ~17--23 & ~~50--70 & ~~60--95 & ~~~40--80 \\ 
$\Gamma_{\rm m}$ & ~~~61 & ~40--70  & ~~90--110 & ~~45--80 & ~~~40--85 \\ 
$\Gamma_{\rm nm}$ & ~~~12 & ~~4--12 & & & ~~~$\sim 20$ \\
\bottomrule 
\end{tabular*} 
\end{center} 

\begin{multicols}{2} 

Results of few-body calculations for the $K^-pp$ system are displayed in 
Table~\ref{tab1}. The marked difference between the `$\bar K NN$ single 
channel' binding energies $B_{K^-pp}$ reflects the difference between the 
input $\bar K N$ amplitudes: the YA $I=0$ single-pole amplitude \cite{YA02} 
resonates at 1405 MeV, whereas the DHW $I=0$ amplitude \cite{DHW08} 
resonates at 1420 MeV (close to the upper of two poles). 
This dependence on the input amplitudes has been verified 
in a recent coupled-channel Faddeev study \cite{SGMR07} 
and in variational calculations~\cite{WG08}. 

A notable feature of the $K^-pp$ coupled-channel calculations 
\cite{SGM07,IS07,WG08} in Table~\ref{tab1} is that the explicit use of 
the $\pi\Sigma N$ channel adds about $20 \pm 5$~MeV to the binding energy 
calculated using effective $\bar K N$ potential within a single-channel 
calculation. This is demonstrated in Fig.~\ref{fig8} by comparing 
corresponding points on the two trajectories shown there. 

Besides present $p(p,K^+)$ measurements \cite{Fab09} at GSI, improving 
on the DISTO $pp \to K^+ \Lambda p$ data, the $K^-pp$ system will be 
explored at J-PARC in the $^3{\rm He}(K^-,n)$ and $d(\pi^+,K^+)$ 
reactions \cite{Nag09}.

\section{Multi-$\bar K$ nucleus quasibound states from RMF calculations} 
\label{RMF}

\begin{center} 
\includegraphics[width=7cm]{12C.eps} 
\figcaption{\label{fig10} $1s$ $K^-$ separation energy $B_{K^-}$ and width 
$\Gamma_{K^-}$ in $^{12}$C, calculated statically (open circles) and 
dynamically (solid circles) as a function of the $\omega KK$ and $\sigma KK$ 
fractional coupling strengths $\alpha_{\omega}$ and $\alpha_{\sigma}$, 
respectively, with $\alpha_{\omega}$ varied in the left panels as indicated 
while $\alpha_{\sigma}=0$, and with $\alpha_{\sigma}$ varied in the right 
panels as indicated while holding $\alpha_{\omega}=1$. The dotted line 
shows $B_{K^-}$ when Im$V_{\bar K}$ is switched off in the dynamical RMF 
calculation \citep{MFG05}.} 
\end{center} 

Dynamical relativistic mean field (RMF) calculations of single-$\bar K$ 
quasibound states yield separation energies in the range 100-150 MeV for 
potentials compatible with $K^-$ atom data \cite{MFG05}. By scanning on the 
$\omega$ vector-field and $\sigma$ scalar-field strengths, these calculations 
also provide a quantitative estimate of the width $\Gamma_{K^-}$ as a function 
of the $K^-$ separation energy $B_{K^-}$. The width comes out larger than 
100 MeV near threshold, decreasing to 50 MeV or slightly more as soon as 
the primary $\bar K N \to \pi\Sigma$ decay mode shuts off 100 MeV below 
threshold \cite{MFG05,GFGM07}. A full systematics for $^{12}$C is shown in 
Fig.~\ref{fig10} which also demonstrates the substantial gain in $B_{K^-}$ 
and $\Gamma_{K^-}$ for $B_{K^-}>100$ MeV when the nuclear core is allowed to 
adjust dynamically in response to the additional mean field generated by 
the $\bar K$ meson. It is worth noting that the resulting nuclear central 
densities do not increase by more than a factor 2 with respect to $\rho_0$, 
and even that is limited to a region of $1-2$~fm about the origin. The figure 
also shows that Im$V_{\bar K}$ may safely be ignored in the calculation of 
separation energies near and above 100 MeV. These results for $B_{K^-}$ and 
$\Gamma_{K^-}$, and for nuclear densities, were shown in the RMF calculations 
of Refs.~\cite{MFG05,GFGM07} to hold over a comprehensive range of nuclei 
from $^{12}$C to $^{208}$Pb. 

\begin{center}
\includegraphics[width=7cm]{multik40ca.eps}
\figcaption{\label{fig11} $K^-$ separation energy $B_{K^-}$, calculated 
in several nuclear RMF models (see inset) for multi-$K^-$ nuclei 
$^{40}{\rm Ca}+\kappa {K^{-}_{1s}}$. The lower (upper) group of curves was 
constrained to produce $B_{K^-}=100~(130)$ MeV for $\kappa=1$ \citep{GFGM08}.} 
\end{center} 

Highlights of multi-$\bar K$ nuclear calculations are summarized below, 
based on recent work by Gazda {\it et al.} \cite{GFGM08,GFGM09}. In order 
to establish correspondence with chiral models, particularly with the 
leading-order TW term, the $\bar K$ coupling constants to the vector meson 
fields were chosen to obey F-type SU(3) symmetry, 
namely $\alpha_{\rm V} \equiv F/(F+D) = 1$: 
\begin{equation} 
\label{eq:SU(3)} 
2g_{\omega KK}=\sqrt{2}\,g_{\phi KK}=2\,g_{\rho KK}=\,g_{\rho \pi\pi}=6.04 \:, 
\end{equation} 
where the value of $g_{\rho \pi\pi}$ is due to the $\rho \to 2\pi$ decay 
width. The value of $g_{\omega KK}^{\rm SU(3)}=3.02$ is lower than any of 
the other choices made in previous works, as detailed in Ref.~\cite{GFGM08}. 
The $\bar K$ coupling constant to the scalar field $\sigma$, $g_{\sigma KK}$, 
was used to fit prescribed values of $B_{K^-}$ in nuclear systems with 
a single $\bar K$ meson. The resulting values of $g_{\sigma KK}$ are also 
lower than those commonly used, e.g. those inspired by quark models. 

$K^{-}_{1s}$ separation energies $B_{K^-}$ in 
multi-$K^-$ nuclei $^{40}{\rm Ca}+\kappa K^-$ are shown in Fig.~\ref{fig11} 
for two choices of $g_{\sigma KK}$, designed within each RMF model to produce 
$B_{K^-}=100$ and $130$ MeV for $\kappa=1$. The difference between the various 
curves, for a given starting value of $B_{K^-}$, originates from the specific 
balance in each one of these RMF models between the vector fields and the 
scalar field. A robust saturation of $B_{K^-}$ with $\kappa$ is observed, 
independently of the applied RMF model, owing to the isoscalar vector-meson 
fields ($\omega,\phi$) which induce repulsion between like $\bar K$ mesons. 
Additional repulsion, in this particular case, is caused by the isovector 
vector-meson field $\rho$. 

The saturation values of $B_{K^-}$ in Fig.~\ref{fig11} are considerably lower 
than what would be required to convert $\Lambda$ hyperons to $\bar K$ mesons 
through strong decays $\Lambda \to p + K^-$, and also $\Xi^-\to\Lambda + K^-$, 
in multi-strange hypernuclei which hence remain the lowest-energy 
configuration for multi-strange systems \cite{SHM93}. This is demonstrated 
in Fig.~\ref{fig12} for various multi-strange hypernuclei constructed under 
realistic assumptions on the meson-field couplings to $\Lambda$ and $\Xi$ 
hyperons \cite{GFGM09}. The figure shows little dependence of $B_{K^-}$ on 
whether or not $\Xi$ hyperons are added to $\Lambda$ hyperons 
(in a $^{208}$Pb core) and very little dependence (in a $^{90}$Zr core) 
on the potential depth assumed for $\Xi$ hyperons in nuclei 
($V^{\rm R}_{\Xi}=-25$ MeV as opposed to the unmarked $V^{\rm R}_{\Xi}=-18$ 
MeV that corresponds relativistically to the nonrelativistic value 
$V^{\rm NR}_{\Xi}=-14$ MeV determined in the BNL-E885 experiment 
\cite{88500}). These RMF calculations across the periodic table provide 
a powerful argument against $\bar K$ condensation under strong-interaction 
equilibrium conditions in terrestial experimentation. It does not apply, 
however, to kaon condensation in neutron stars, where equilibrium 
configurations are determined by weak interactions. 

\begin{center}
\includegraphics[width=7cm]{multikLXi.eps}
\figcaption{\label{fig12} RMF calculations of multi-$\bar K$ quasibound 
states as a function of the number $\kappa$ of $\bar K$ mesons in 
multi-strange nuclei \citep{GFGM09}.} 
\end{center} 

\acknowledgments{I am indebted to my collaborators Ale\v{s} Ciepl\'{y}, Eli 
Friedman, Daniel Gazda and Ji\v{r}\'{i} Mare\v{s} for longtime fruitful 
cooperation, and to Wolfram Weise for many instructive discussions. Special 
thanks are due to the organizers of QNP09 for their kind hospitality. } 

\end{multicols} 

\vspace{-2mm}
\centerline{\rule{80mm}{0.1pt}}
\vspace{2mm}

\begin{multicols}{2}

\end{multicols}

\clearpage

\end{document}